\documentclass[a4paper,11pt]{article}

\pdfoutput=1
\usepackage{subfig}
\usepackage{booktabs}

\usepackage{lineno}
\usepackage[colorlinks=false]{hyperref}
\usepackage{bm}
\usepackage{graphicx}
\usepackage{amssymb}
\usepackage{amsmath}
\usepackage{tabularx}
\usepackage{natbib}
\usepackage{floatrow}
\usepackage{caption}
\usepackage{amsmath}
\usepackage{amsfonts}
\usepackage{amssymb,multirow,enumerate,threeparttable}
\usepackage[decimalsymbol=fullstop,digitsep=space,sepfour=true]{siunitx}
\newcommand{\cm}{{\mathcal M}}

\newcommand*{\abs}[1]{\left|{#1}\right|}
\newcommand*{\bracket}[1]{\left({#1}\right) } 

\usepackage{authblk}

\usepackage[left=15mm,right=15mm,top=1.5cm,bottom=1.5cm,includeheadfoot]{geometry}
\setcitestyle{square,numbers}

\begin{document}

\title{Discrepancies of measured SAR between traditional and fast measuring systems}


\author[1]{Zicheng Liu}
\author[2]{Djamel Allal}
\author[3]{Maurice Cox}
\author[1]{Joe Wiart}

\affil[1]{\scriptsize Chaire C2M, LTCI, T\'el\'ecom Paris, 91120 Palaiseau, France}
\affil[2]{\scriptsize Laboratoire National de Métrologie et d’Essais, 78197 Trappes, France}
\affil[3]{\scriptsize National Physical Laboratory, Teddington TW11 0LW, UK}

\maketitle

\abstract{
Human exposure to mobile devices is traditionally measured by a system in which the human body (or head) is modelled by a phantom and the energy absorbed from the device is estimated based on the electric fields measured with a single probe. Such a system suffers from low efficiency due to repeated volumetric scanning within the phantom needed to capture the absorbed energy throughout the volume. To speed up the measurement, fast SAR (specific absorption rate) measuring systems have been developed. However, discrepancies of measured results are observed between traditional and fast measuring systems. In this paper, the discrepancies in terms of post-processing procedures after the measurement of electric field (or its amplitude) are investigated. Here, the concerned fast measuring system estimates SAR based on the reconstructed field of the region of interest while the amplitude and phase of electric field are measured on a single plane with a probe array. The numerical results presented indicate that the fast SAR measuring system has the potential to yield more accurate estimations than the traditional system, but no conclusion can be made on which kind of system is superior without knowledge of the field-reconstruction algorithms and the emitting source.
\\

\noindent
\begin{tabular}{ll}
\textbf{Keywords:} &Specific absorption rate, fast SAR measurement, field reconstruction, plane-wave \\&expansion, traditional SAR measurement, measurement discrepancy, uncertainty analysis
\end{tabular}
}

\section{Introduction}
Human exposure \cite{kshetrimayum2008mobile} has drawn much public attention recently due to the wide usage of wireless communication equipment. Ensuring the absorbed energy lies within the safe range \cite{van2011research} requires the accurate quantification of the so-called specific absorption rate (SAR). Efforts towards such quantification have been made by researchers both in simulation tools \cite{wiart2016radio,chiaramello2019radio,christ2009virtual} and measuring systems \cite{friden2006quick,pfeifer2018total,sasaki2019error,IECSARStandard}. For simulation techniques, when the distance is large (relative to the wavelength of the electric field), the front of incident wave upon the body can be treated as planar \cite{conil2011influence,hirata2007dominant}, while near fields need to be analyzed with a high-resolution human module with a numerical code (FDTD \cite{martinez2009fdtd} or FEM \cite{meyer2003human}) when the human body is close to sources \cite{chiaramello2017assessment,lacroux2008specific,chiaramello2017stochastic}. In measurements, the exposure to base stations is usually estimated by measuring the intensity of the electric field at various locations and through the construction of a path loss model it is possible to have a statistical view \cite{chobineh2019statistical}. The exposure to user equipment is traditionally measured by moving a probe inside a liquid-filled phantom, which simulates the composition of human body or head, to estimate the whole-body and/or local SAR \cite{IECSARStandard}.  

The traditional measuring system \cite{IECSARStandard} when applied to user equipment is the concern of this paper. The peak spatial-average (\SI{1}{\gram} or \SI{10}{\gram}) SAR is often of interest in practice.
The probe carries out the so-called area scan (two-dimensional scanning over a coarse grid to find the location of maximum SAR) and zoom scan (three-dimensional scanning with a finer grid to determine the peak mass-averaged SAR using interpolation and extrapolation techniques). A complete measurement consumes tens of minutes while more than 100 measurements have to be performed to check full compliance (with different frequency bands, working modes and device positions) of a product. Considering that with today's massive production the compliance of millions (even billions) of products needs to be assessed, the efficiency of such a measuring system is intolerable and in great demand to be improved.  

Research has been carried out to speed up the process by reducing the number of measurement points and deducing the peak spatial-average SAR based on parametric models \cite{merckel2004parametric} or empirical observations \cite{kanda2004faster}. However, since the method is model-dependent, the estimation accuracy for newly emerged devices (e.g., equipped with MIMO terminals) is not guaranteed. The estimation approach based on the technique of plane-wave expansion \cite{kong1986electromagnetic} has been proposed in which only the electromagnetic properties of the medium are required. Such an approach has recently been applied to the computation of power density of millimetre waves \cite{sasaki2019error}.

As described above, various approaches exist to obtain improvements in SAR measurement. The problem arises when discrepant results are observed from fast measuring systems that are developed by different manufacturers \cite{compSAR2019}. Due to reasons of commercial security, it is hard to justify which system is more accurate than others and also not easy to conclude that the fast measuring system is biased when the generated results challenge the traditional system, which is usually considered as the reference.

Efforts are made in this paper to reveal the reason for estimation discrepancies and try to answer the following questions:
\begin{itemize}
	\item	Why do discrepancies appear for the estimation of SAR by different fast measuring systems?
	\item	Can we say fast measuring systems generate biased estimations if they differ appreciably from the traditional SAR measuring system? 
	\item	Which of the traditional measuring system and the fast  measuring system is the more accurate?
\end{itemize}
Despite various methodologies for the fast SAR measuring system, the concerned system measures the electric field inside the phantom by a vector-probe on a plane. Then, the field in other positions of interest is reconstructed by algorithms (e.g.\ plane-wave expansion), which can be carried out with high computational efficiency. The study is based on analytical functions, which simulate the wave propagation of electric field inside the phantom. To avoid too complex an analysis, the estimation discrepancies due to the post-processing procedures are mainly investigated and factors like scattering from the phantom shell will not be considered. 

The remaining part of this paper is organized as follows. Section \ref{sec:traditionalSAR} and \ref{sec:fastSAR} introduce the methodology of SAR estimation in the traditional system and the concerned fast measuring system, respectively. Numerical results with the flat phantom and $11$ emitting sources in Section \ref{sec:NumericalResults} indicate the accuracy and stability of the fast estimation and present the estimation discrepancies between two kinds of systems. Conclusions are made in Section \ref{sec:conclusion}.

\section{Traditional SAR measuring system}
\label{sec:traditionalSAR}

\begin{figure}[!ht]
	\centering
	\includegraphics[width=0.38\linewidth]{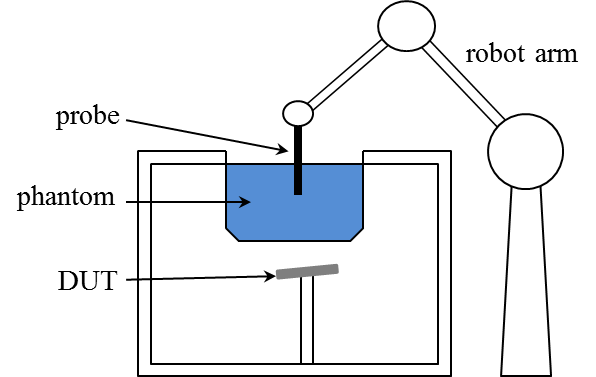}
	\caption{Sketch of traditional SAR measuring system.}
	\label{fig:sketchTraditionalSAR}
\end{figure}

The definition of the spatial-average specific absorption rate (SAR) is 
\begin{equation}
\text{SAR}=\frac{1}{V}\int_{\mathbb{V}}\frac{|\mathbf{E}(\mathbf{r})|^2\sigma(\mathbf{r})}{\rho(\mathbf{r})}\mathrm{d}\mathbf{r},
\label{eq:defSAR}
\end{equation}
where $V$ is the volume of the region of interest $\mathbb{V}$, $|\mathbf{E}(\mathbf{r})|$ is the root-mean-square of electric field at the location $\mathbf{r}$, and $\sigma$ and $\rho$ denote conductivity and density, respectively. $\mathbb{V}$ is a cube with side length \SI{1}{\mm} for \SI{1}{\gram} SAR and \SI{21.5}{\mm} for \SI{10}{\gram} SAR when $\rho = \SI{1}{\gram/\cm^3}$. The value of $\sigma$ depends on the wave frequency and the reference value can be found in the literature \cite{drossos2000dependence,fields1997evaluating}. As seen from \eqref{eq:defSAR}, the phase of the electric field is not required and thus only the amplitude is measured by the probe in the traditional measuring system, the composition of which is sketched in Figure \ref{fig:sketchTraditionalSAR}, where a probe is moved by a robot arm to measure the amplitude of electric fields inside a phantom due to the emitting device under test. The integration in \eqref{eq:defSAR} is carried out numerically (e.g., by the trapezoidal rule) and the amplitude of the electric field is required at dense sampling points inside $\mathbb{V}$. 

Rather than making intensive measurements inside the whole phantom, area scan and zoom scan are performed sequentially and followed by interpolations and extrapolations to provide the amplitude of the electric fields at the desired locations. The procedure is summarized below. 
\begin{enumerate}
	\item	{\em Area scan}: measure fields according to a two-dimensional coarse grid, the distance of which to the phantom surface is fixed, to locate the local maxima of the amplitude of electric fields. 
	\item	{\em Zoom scan}: a three-dimensional scanning within cubes centered at the location of local maxima, the grid step being smaller than that in the area scan.
	\item	{\em Interpolation and extrapolation}: linear interpolation and cubic spline interpolation (and extrapolation) are used as necessary to deduce the amplitude at the points in a finer grid.
	\item   {\em Peak spatial-average SAR}: obtained by performing numerically the integration in \eqref{eq:defSAR} based on the interpolated and extrapolated amplitude.
\end{enumerate}
Documentary standards are available for the specific requirements on the above measurements and post-processing. Here, the standard \cite{IECSARStandard} is followed, which includes the requirements in Table \ref{tab:configTradSAR}. For the interpolation and extrapolation, no specific algorithms are required or recommended. Here, the method of linear and cubic spline interpolation are applied. Note that the SAR drift is also tested for practical instruments, but not considered during the simulation.

\begin{table}[!ht]
	\centering
	\begin{threeparttable}
		\caption{Configurations of traditional SAR measuring system.}
		\label{tab:configTradSAR}
		\begin{tabular}{cp{4.9cm}p{7.3cm}}
			\toprule
			\multirow{3}{*}{Area scan} & maximum grid spacing & \SI{20}{\mm} if $f<$ \SI{3}{\GHz} and $60/f$\,\si{\mm} otherwise \\
			\cline{2-3}
			& maximum distance between probe and surface of phantom & \multirow{2}{*}{\SI{5}{\mm} if $f<$ \SI{3}{\GHz} and $\delta\ln 2/2$\,\si{\mm} otherwise}\\
			\midrule
			\multirow{5}{*}{Zoom scan} & horizontal grid spacing & $\leq \min\{24/f, 8\}$\,\si{\mm} \\
			\cline{2-3}
			& \multirow{2}{*}{minimum scan size} & $\SI{30}{\mm} \times \SI{30}{\mm} \times \SI{30}{\mm}$ if $f <$ \SI{3}{\GHz} and $\SI{22}{\mm} \times \SI{22}{\mm} \times \SI{22}{\mm}$ otherwise \\
			\cline{2-3}
			& maximum distance between probe and surface of phantom & \multirow{2}{*}{\SI{5}{\mm} if $f <$ \SI{3}{\GHz} and $\delta\ln 2/2$\,\si{\mm} otherwise} \\
			\bottomrule
		\end{tabular}
	\end{threeparttable}
		\small
		\item ``$\ln$" denotes natural logarithm, $f$ wave frequency in \si{\GHz}, $\delta$ plane-wave skin depth 
\end{table}

\section{Fast SAR measuring system based on field reconstruction}
\label{sec:fastSAR}
\begin{figure}[!ht]
	\centering
	\includegraphics[width=0.38\linewidth]{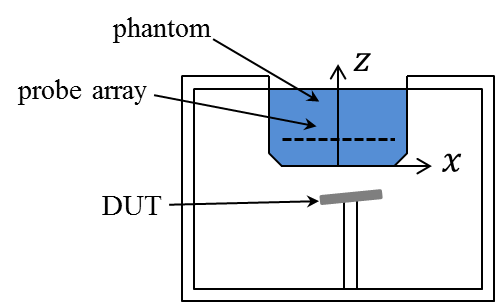}
	\caption{Sketch of the concerned fast SAR measuring system.}
	\label{fig:sketchFastSAR}
\end{figure}

Rather than a single probe, a probe array is used in the fast SAR measuring system, which is sketched in Figure~\ref{fig:sketchFastSAR}. The probes measure the amplitude and phase of electric fields on a single plane. The solution to fields at other positions of interest is obtained by a field-reconstruction algorithm. Note that the algorithms used in commercial products are usually inaccessible for reasons of security protection. Here, the technique of plane-wave expansion (PWE), which is popularly utilized when measurements are taken on plane(s), is applied. To study the effects of different field-reconstruction algorithms, the PWE method, which is found inaccurate for some cases, is performed with different settings.

\subsection{Plane-wave expansion (PWE)}
\label{sec:PWE}

Based on the PWE theory, the electric field can be represented as an integral of planar waves,
\begin{equation}
E(x,y,z) = \frac{1}{4\pi^2}\int_{-\infty}^{\infty}\int_{-\infty}^{\infty} E(k_x,k_y)e^{i(k_xx+k_yy-k_zz)}\mathrm{d}k_x\mathrm{d}k_y,
\label{eq:pwe}
\end{equation}
where $(x,y,z)$ is the Cartesian coordinates of the observation point, $k_x$, $k_y$ and $k_z$ are respectively the $x$, $y$, $z$ component of the wavenumber of the liquid $k$ with the identity $k_x^2+k_y^2+k_z^2=k^2$. The imaginary part of $k_z$ is non-negative so that the Sommerfeld radiation condition \cite{sommerfeld1949partial} is satisfied. Note that \eqref{eq:pwe} is valid for the $x$, $y$ and $z$ component of electric fields and the three components can be independently treated. Thus, the following derivations apply to each component.

From \eqref{eq:pwe}, we see that $E(x,y,z)$ is a two-dimensional inverse Fourier transform of $E(k_x,k_y)e^{-ik_zz}$, i.e., $E(x,y,z)$ can be reconstructed if the corresponding spectrum $E(k_x,k_y)e^{-ik_zz}$ is known. Set $E(k_x,k_y,z) = E(k_x,k_y)e^{-ik_zz}$. It follows that 
\begin{equation}
E(x,y,z_{\text{rec}}) = \frac{1}{4\pi^2}\int_{-\infty}^{\infty}\int_{-\infty}^{\infty} E(k_x,k_y,z_{\text{mea}})e^{-ik_z(z_{\text{rec}}-z_{\text{mea}})}e^{i(k_xx+k_yy)}\mathrm{d}k_x\mathrm{d}k_y,
\label{eq:pwe_mea}
\end{equation}
where $z_{\text{mea}}$ and $z_{\text{rec}}$ denote the $z$-coordinate of the measurement and reconstruction plane, respectively, and the multiplicand $e^{-ik_z(z_{\text{rec}}-z_{\text{mea}})}$ is the so called spectral propagator. From \eqref{eq:pwe}, we see $E(k_x,k_y,z_{\text{mea}})$ is the two-dimensional Fourier transform of $E(x,y,z_{\text{mea}})$, i.e.,  
\begin{equation}
E(k_x,k_y,z_{\text{mea}}) = \int_{-\infty}^{\infty}\int_{-\infty}^{\infty}E(x,y,z_{\text{mea}}) e^{-i(k_xx+k_yy)}\mathrm{d}x\mathrm{d}y,
\label{eq:FT_mea}
\end{equation}
which is to be estimated based on the measured fields. 

Electric fields are measured on a plane with probes uniformly distributed in a specific domain, whose centre is usually within the main lobe of the radiated field. Denote by $\Delta x$ and $\Delta y$ the minimum interval between probes along the $x$ and $y$ axis, respectively. The coordinates of the probes are $(x_p^{n_x},y_p^{n_y})=(x_p^\mathrm{c},y_p^\mathrm{c})+n_x\hat{x}+n_y\hat{y}$, where $(x_p^\mathrm{c},y_p^\mathrm{c})$ is the centre of the measurement domain, $n_x$, $n_y$ are integers in the intervals $[-(N_x-1)/2,(N_x-1)/2]$ and $[-(N_y-1)/2,(N_y-1)/2]$, respectively. $N_xN_y$ is the number of probes and assumed odd. $\widehat{x}$ and $\widehat{y}$ are unit vectors along the $x$ and $y$ axis, respectively.

Assume that the amplitude of electric fields outside the measurement domain is quite small.
The sampling density satisfies the Nyquist sampling theorem, i.e., $2\pi/\Delta x \ge k_x^{\text{max}}$, $2\pi/\Delta y \ge k_y^{\text{max}}$, and the spectrum $E(k_x,k_y)$ is approximately zero outside the intervals $-k_x^{\text{max}}/2\le k_x \le k_x^{\text{max}}/2$, $-k_y^{\text{max}}/2\le k_y \le k_y^{\text{max}}/2$. Then \eqref{eq:FT_mea} can be well approximated by the discrete form 
\begin{equation}
E(k_x^{m_x},k_y^{m_y},z_{\text{mea}}) = \sum_{n_x=-(N_x-1)/2}^{(N_x-1)/2}\,\,\,\sum_{n_y=-(N_y-1)/2}^{(N_y-1)/2}  E(x^{n_x},y^{n_y},z_{\text{mea}})e^{-i(k_x^{m_x}x^{n_x}+k_y^{m_y}y^{n_y})},
\label{eq:DFT}
\end{equation}
where $k_x^{m_x}=m_xk_x^{\text{max}}/M_x$, $m_x=-(M_x-1)/2,\ldots,(M_x-1)/2$ and $k_y^{m_y}=m_yk_x^{\text{max}}/M_y$, $m_y=-(M_y-1)/2,\ldots,(M_y-1)/2$, $M_x$, $M_y$ being the number of sampled spatial-frequency points, which are set as odd numbers. Setting the centre $(x_p^c,y_p^c)$ of the measurement domain as the origin of the $x$ and $y$ axis, i.e., $x_p^c =0$, $y_p^c =0$, \eqref{eq:DFT} is rewritten as the standard form of the discrete Fourier transform (DFT),
\begin{equation}
E(m_x,m_y,z_{\text{mea}}) = \sum_{n_x^\prime=0}^{N_x-1}\sum_{n_y^\prime=0}^{N_y-1}  E(n_x,n_y,z_{\text{mea}})e^{-i2\pi( m_xn_x^\prime/M_x+m_yn_y^\prime/M_y)},
\label{eq:standardDFT}
\end{equation} 
with the notation $n_x^\prime = n_x+(N_x-1)/2$, $n_y^\prime = n_y+(N_y-1)/2$. 

Then $E(k_x,k_y,z_{\text{rec}})$ is computed as $E(k_x,k_y,z_{\text{mea}})e^{-ik_z(z_{\text{rec}}-z_{\text{mea}})}$ and the field solution in the spatial domain is obtained after the inverse DFT, i.e.,
\begin{equation}
E(n_x,n_y,z_{\text{rec}}) = \frac{1}{M_xM_y}\sum_{m_x=0}^{M_x-1}\sum_{m_y=0}^{M_y-1} E(m_x,m_y,z_{\text{rec}}) e^{i2\pi( m_xn_x^\prime/M_x+m_yn_y^\prime/M_y)}.
\label{eq:standard_invDFT}
\end{equation}  
Set $\mathbf{F}=[e^{-i2\pi( m_xn_x^\prime/M_x+m_yn_y^\prime/M_y)}]$, $\mathbf{invF} = [e^{i2\pi( m_xn_x^\prime/M_x+m_yn_y^\prime/M_y)}/(M_xM_y)]$,  and $\mathbf{P}(z_{\text{rec}}-z_{\text{mea}}) = \text{diag}\{e^{-ik_z(z_{\text{rec}}-z_{\text{mea}})}\}$. The above reconstruction procedures are expressed in matrix form as
\begin{equation}
\mathbf{E}_{\text{rec}} = \mathbf{invF}\cdot\mathbf{P}\cdot\mathbf{F}\cdot\mathbf{E}_{\text{mea}},
\label{eq:matrixFullFourier}
\end{equation}
where $\mathbf{E}_{\text{rec}}$, $\mathbf{E}_{\text{mea}}$ are column vectors composed of $N_xN_y$ sampled electric fields. In summary, the solution to the field on the reconstruction plane is an inverse FT (denoted by $\mathbf{invF}$) of the spectrum, which is obtained by multiplying the spectrum of measured fields $\mathbf{E}_{\text{mea}}$ by the propagator $\mathbf{F}$.

The PWE approach suffers from being an ill-conditioned problem \cite{tikhonov1963solution} when reconstructing high spatial-frequency (mentioned as frequency in the remaining part) components. When $k_z$ has a large imaginary part, i.e., $k_x^2+k_y^2\gg k^2$, the value of the propagator $e^{-ik_z(z_{\text{rec}}-z_{\text{mea}})}$ would be very large when $z_{\text{rec}}>z_{\text{mea}}$. As a result, the effects of approximation errors (e.g., due to invalid assumptions for \eqref{eq:DFT}) or measurement noises are amplified by the propagator and the estimation would be highly biased and with a large variance.  

Considering that the energy of high-frequency components is usually small, a stable reconstruction is often reached, without losing much accuracy, by only considering the spectrum at low frequencies. With lossless cases, only propagating plane-wave components, i.e., $k_x^2+k_y^2<k^2$, are considered. In the concerned cases, since the equivalent liquid is lossy (wavenumber $k$ is complex), the integrand stands for evanescent waves. However, a similar constraint  $k_x^2+k_y^2\le |k|^2$ is applied.

Note that in practice, only the $x$ and $y$ component of the electric field, denoted by $E_x$ and $E_y$ respectively, are required to be measured or reconstructed, since the $z$ component can be obtained based on the identity
\begin{equation}
E_x(k_x,k_y)k_x + E_y(k_x,k_y)k_y+E_z(k_x,k_y)k_z = 0.
\label{eq:identityElectricField}
\end{equation}

\subsection{Field reconstruction making use of more high-frequency components}
\label{sec:moreHighSpectrum}

When the energy of high-frequency components is small and can be neglected, reconstructing only the low-frequency spectrum is reasonable. Otherwise, since a part of the energy is not taken into account in the algorithm, the estimated SAR value tends to be underestimated, as shown by the numerical results in Section \ref{sec:NumericalResults}. Thus, high-frequency components need to be reconstructed. Replacing the constraint of the spectrum by $\sqrt{k_x^2+k_y^2}<\varepsilon$, a higher-frequency spectrum would be considered as $\varepsilon$ increases. For the commonly utilized approach in Section \ref{sec:PWE}, the threshold $\varepsilon$ is set as $|k|$. To quantify the additional spectrum considered, $\varepsilon$ is expressed by $|k|+(\sqrt{k_{x,\max}^2+k_{y,\max}^2}-|k|)\delta$. The parameter $\delta\in[0,1]$ quantifies how much high-frequency spectrum is reconstructed. For the traditional PWE method, $\delta=0$. The reconstruction of the complete spectrum is achieved by setting $\delta=1$.

\section{Numerical results}
\label{sec:NumericalResults}

The hyperparameters in the described PWE method include the number of frequency points $M_x$, $M_y$ and $\delta$. $M_x$ and $M_y$ are set as $2N_x+1$ and $2N_y+1$, respectively, while no significant effect is observed with larger values. Different values are given to $\delta$ to study the effects of reconstructing the high-frequency spectrum. 

\subsection{Configurations}
The studied fast measuring system is sketched in Figure~\ref{fig:sketchFastSAR}, where the device under test is placed \SI{5}{\mm} away from the flat phantom and the vector-probe array at the plane with $z = \SI{19.25}{\mm}$. 
An array of $29\times29$ probes is uniformly distributed in the surface defined by $\SI{-10}{\cm} \leq x, y \leq \SI{10}{\cm}$ with interval \SI{7}{\mm}. The measured electric field inside the phantom is obtained based on the analytical function
\begin{equation}
E(\mathbf{r})=\sum_{i=1}^{80}\sum_{j=1}^{80}\bracket{\nabla^2G_k(\mathbf{r}-\mathbf{d}_{i,j})\mathbf{p}_{i,j}+k^2G_k(\mathbf{r}-\mathbf{d}_{i,j})\mathbf{p}_{i,j}}.
\label{eq:analyticalFun}
\end{equation}
The scatterings by different emitting sources are simulated by varying the setting of dipole position $\mathbf{d}_{i,j}\in\mathbb{R}^3$ and moment $\mathbf{p}_{i,j}\in\mathbb{R}^3$. $\mathbf{r}$ denotes the position of the observation point, $G_k(\mathbf{r})=e^{ik\abs{\mathbf{r}}}/4\pi\abs{\mathbf{r}}$ and $k$ is the wavenumber of the phantom. The $11$ cases tested are with the flat phantom and configured according to Table \ref{tab:parameters}. Note that the reference value of \SI{1}{\gram} spatial-average SAR (mentioned as \SI{1}{\gram} SAR in the following part) equals 1 for all cases.
\begin{table}[!h]
	\centering
	\begin{tabular}{cccccccccccc}
		\hline
		Index & 1 & 2 & 3 & 4 & 5 & 6 & 7 & 8 & 9 & 10 & 11\\
		\hline
		f (MHz) & 850 & 1800 & 1900 & 2450 & 5500 & 5800 & 750 & 1950 & 750 & 835 & 1750\\
		\hline
		$\epsilon_r$ & 42.23 & 40.45 & 40.28 & 39.37 & 33.30 & 32.64 & 42.47 & 40.20 & 42.47 & 42.26 & 40.53\\
		\hline
		$\sigma$ (S/m) & 0.89 & 1.39 & 1.45 & 1.87 & 5.18 & 5.55 & 0.85 & 1.49 & 0.85 & 0.88 & 1.35\\
		\hline
		10g SAR & 0.58 & 0.48 & 0.48 & 0.43 & 0.29 & 0.28 & 0.28 & 0.41 & 0.66 & 0.65 & 0.52\\
		\hline
	\end{tabular}
	\caption{Physical parameters and the reference value of \SI{10}{\gram} spatial-average SAR.}
	\label{tab:parameters}
\end{table}	

\subsection{Verification of post-processing procedures}
\label{sec:verification}
\begin{figure}[!h]
	\centering
	\includegraphics[width=0.8\linewidth]{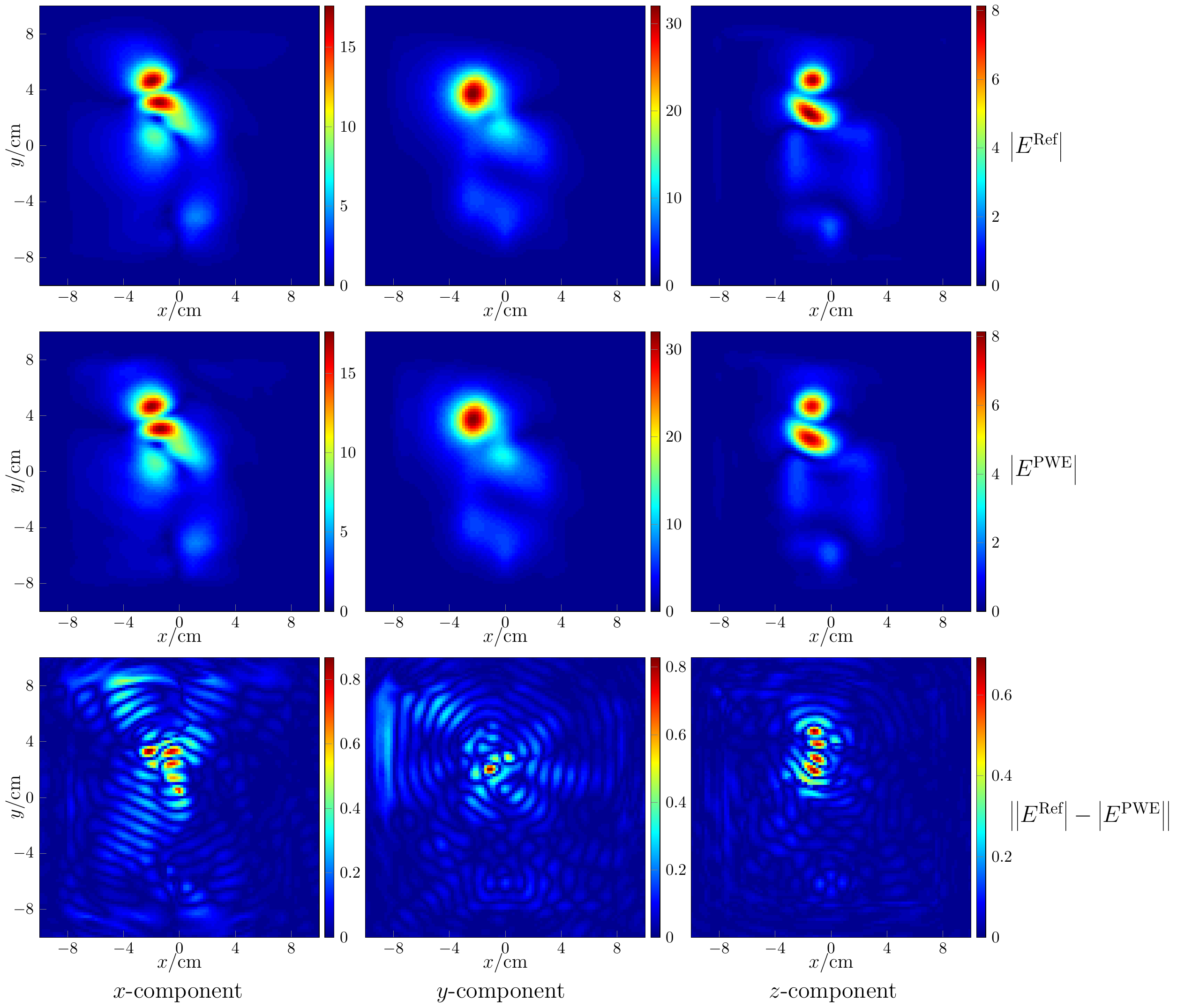}
	\caption{Field reconstruction with respect to the 4th case. $E^\text{PWE}$ and $E^\text{Ref}$ denote the reconstructed field and the reference field, respectively.}
	\label{fig:verifyFieldRecon}
\end{figure}

The post-processing includes the field reconstruction of regions of interest and the followed computation of peak spatial-average SAR. The former is verified by observing the reconstructed field on the plane of $z = \SI{0}{\mm}$ and comparing with the reference as shown in Figure 3 for the 4th case, where the maps of field intensity in the first row are the reference, the second row gives the reconstructed results, and the last row shows the absolute error of reconstruction. The resolution of the map is \SI{2}{\mm}. As seen, the amplitude of the reconstructed fields is a little different from the reference, the absolute errors being small. Setting the 
resolution as \SI{1}{\mm}, with the reconstructed field of the cube bounded by $\SI{-10}{\cm} \leq x, y \leq \SI{10}{\cm}$, $\SI{0}{\cm} \le z \le \SI{3}{\cm}$, the computed peak spatial-average SAR equals $0.993$ for \SI{1}{\gram} and $0.427$ for \SI{10}{\gram}, which are quite close to the reference value $1$ and $0.431$. 

\begin{figure}[!h]
	\centering
	\includegraphics[width=0.4\linewidth]{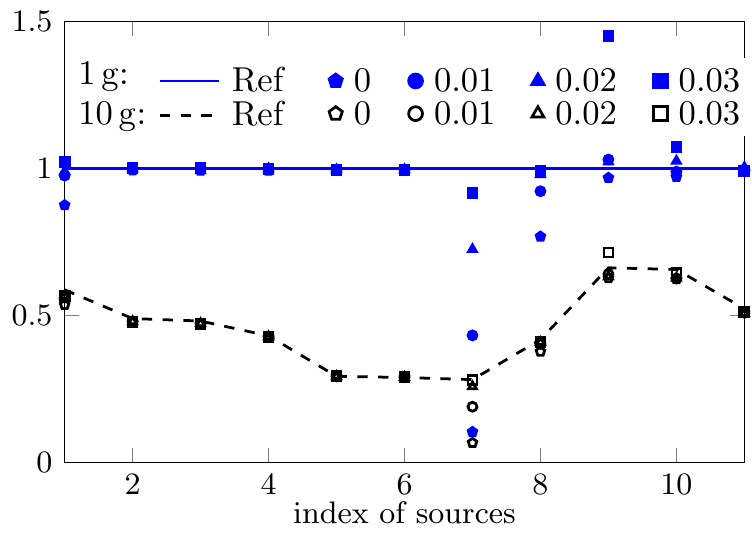}
	\caption{Estimation of \SI{1}{\gram} and \SI{10}{\gram} spatial-averaged SAR based PWE field-reconstruction method by assigning four different values to $\delta$.}
	\label{fig:valSARCompRef}
\end{figure}

The computed \SI{1}{\gram} and \SI{10}{\gram} SAR of the $11$ cases based on the reconstructed field are shown in Figure~\ref{fig:valSARCompRef}. For most cases, with the commonly used PWE approach, i.e., setting $\delta = {0}$, the SAR value can be accurately estimated. However, e.g., with respect to the 1st, 7th, and 8th case, increasing the value of $\delta$ significantly improves the estimation accuracy since a larger part of high spatial-frequency spectrum is considered. Note that the value of $\delta$ should not be too large due to the involved ill-conditioned problem. For instance, when $\delta$ is increased to 0.03, the estimated SAR value changes from closely approximated solutions to a highly biased one. An insight into the ill-conditioned problem is given through the investigation of the field reconstruction problem with respect to the 7th case and the uncertainty analysis.  

\subsection{Problem in field reconstructions}	
\label{sec:problem}

\begin{figure}[!h]
	\centering
	\subfloat[Amplitude of electric field]{%
		\includegraphics[clip,width=\columnwidth]{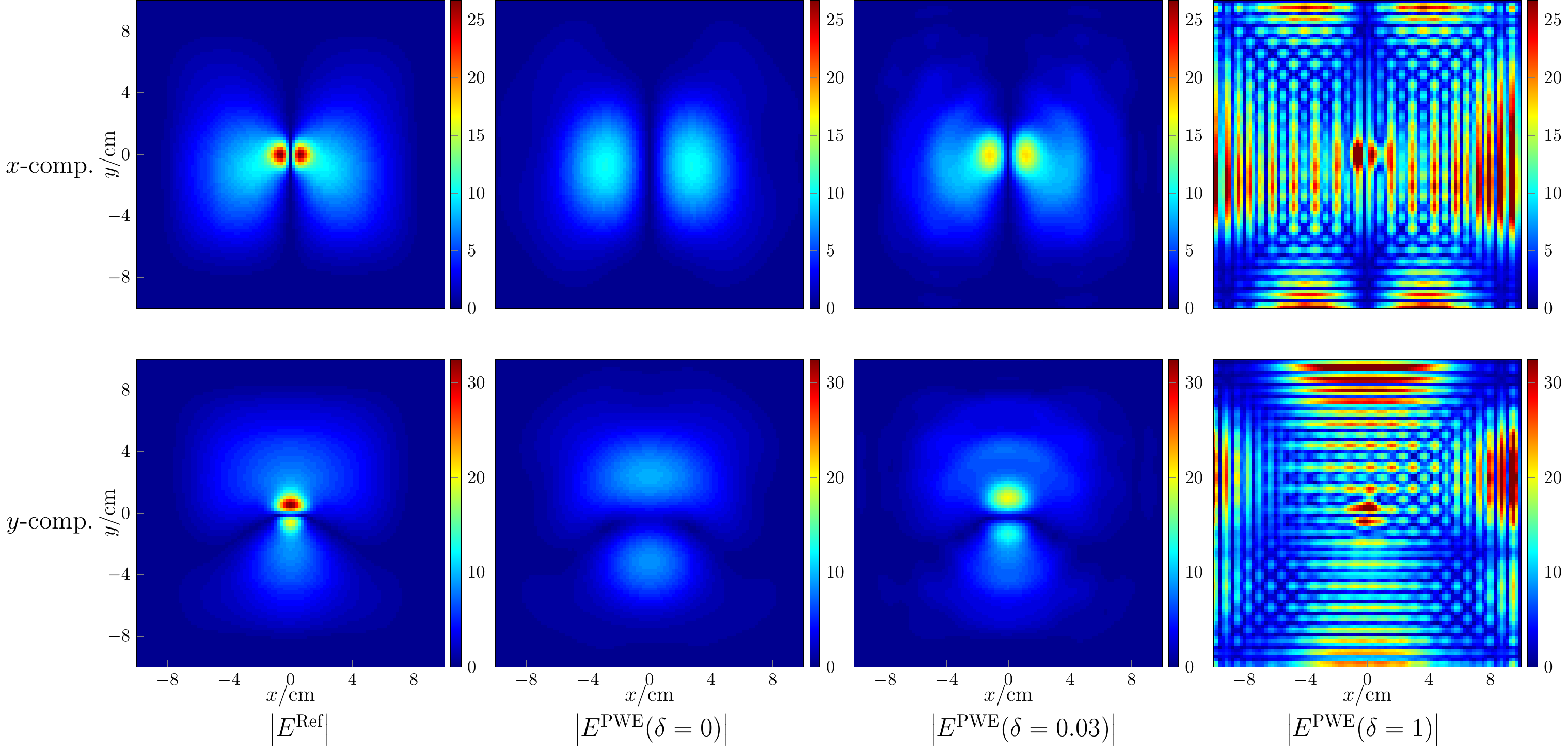}\label{fig:electricField7Source}
	}
	
	\subfloat[Amplitude of spectrum]{%
		\includegraphics[clip,width=\columnwidth]{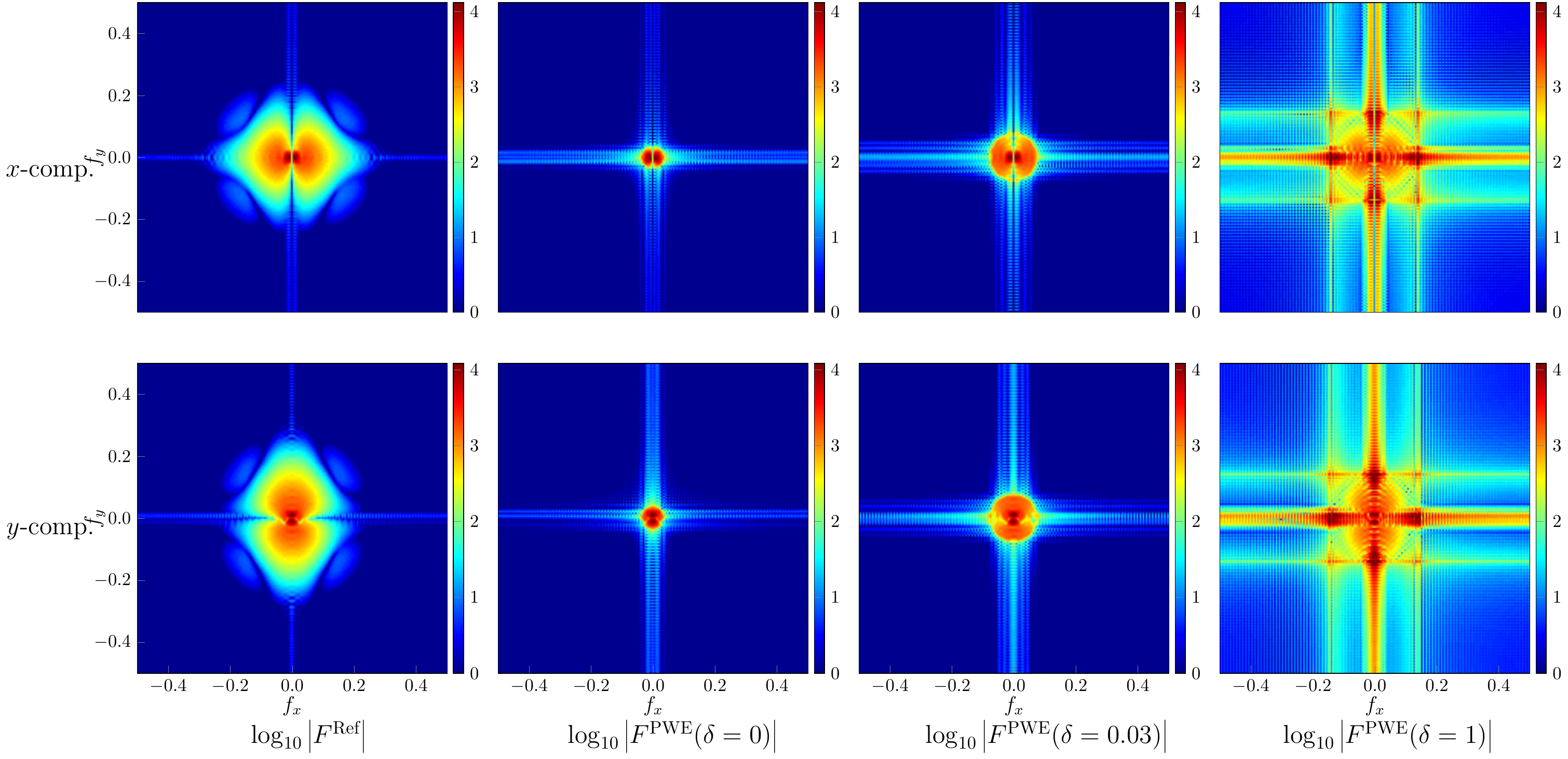}%
		\label{fig:spectrum7Source}}
	
	\caption{Field reconstruction with respect to the 7th case, $|E|$ denotes the amplitude of electric field, $|F|$ the amplitude of spectrum, and the superscript ``PWE", ``Ref" indicate the reconstructed field and the reference field, respectively.}
\end{figure}

While the development of advanced field-reconstruction algorithms is not the concern of this paper, the challenges encountered in the field reconstruction can be presented. With respect to the 7th case, where the value of SAR is underestimated, the amplitude of the reconstructed $x,y$ component of electric fields is shown in Figure~\ref{fig:electricField7Source}. The colour limits are set to be the same for each component. Note that since the $z$ component is determined by the $x,y$ counterparts, only the $x$ and $y$ components of the electric field are shown. As observed, the electric fields with a large amplitude are distributed intensively in the spatial domain. Consequently, as shown by Figure~\ref{fig:spectrum7Source}, the energy of the spectrum spreads widely and the high-frequency spectrum cannot be neglected. If following the commonly utilized approach and only reconstructing the spectrum constrained by $k_x^2+k_y^2\le |k|^2$, the smooth region of the electric field is well reconstructed but the small regions, which are centered at the peak values and influential to the estimation of the peak spatial-average SAR, are poorly reconstructed. Increasing $\delta$ by 0.03 to make more use of the spectrum, the reconstruction accuracy is improved. However, setting $\delta$ to 1 to consider the complete spectrum, very poor reconstructions are observed. From the corresponding spectrum, it is seen that the low-frequency spectrum is well retrieved but the high-frequency part suffers from large deviations. That is due to the ill-conditioning mentioned in Section \ref{sec:PWE}. The effects of approximation errors in the PWE approach (e.g., due to insufficiently small sampling density) are amplified after the back-propagation stage of the reconstruction algorithm. 

\subsection{Uncertainty of factors}
\label{sec:uncertainty}

Assume the concerned factors follow the distributions described in Table \ref{tab:pdf}, where the superscript ``Ref" denotes the reference or recorded value. The mutual inductance among probes yields the so-called coupling effects. In practice, since the composition of probes may not be exactly the same, the coupling effects can differ for different couples of probes and are hard to be analyzed due to many involved coupling coefficients. Here, the coupling effects of each probe are analyzed by only taking into account the mutual inductance from its 8 neighbour probes. As a result, the measured noisy field is obtained by 
\begin{equation}
\mathbf{E}(\mathbf{r}) = \mathbf{E}^\text{Ref}(\mathbf{r}) + \sum_{p_x=-1}^{1}\sum_{p_y=-1}^{1}\begin{bmatrix}c_{p_x,p_y}^{x,x} & c_{p_x,p_y}^{x,y} \\ c_{p_x,p_y}^{y,x} & c_{p_x,p_y}^{y,y}\end{bmatrix}\mathbf{E}_{p_x,p_y}^\text{Ref}(\mathbf{r}),
\label{eq:coupling}
\end{equation} 
where the subscripts $p_x$ and $p_y$ indicate the position of the neighboured probes while $\mathbf{E}_{0,0}^\text{Ref}$ denote the electric field measured by the concerned probe. Note that the reference of the measured field is computed by \eqref{eq:analyticalFun} when the uncertainty of the probe position and the phantom electromagnetic properties has been taken into account. Since  the $z$-component is usually computed (rather than measured) according to \eqref{eq:identityElectricField}, the electric field in \eqref{eq:coupling} is a column vector composed of the $x$ and $y$ component.

\begin{table*}[!ht]
	\centering
	\caption{Description and distribution of input variables. $\mathcal{U}(a, b)$ denotes the uniform distribution with limits $a$ and $b$, and $\mathcal{N}(\mu, \tau)$ denotes the normal distribution with mean $\mu$ and standard deviation $\tau$.}
	\label{tab:pdf}
	\begin{tabular}{lllll}
		\toprule
		Variable & Description & Distribution\\
		\midrule
		$x_p,y_p,z_p$ (mm) & Cartesian coordinates of the probe position & $a_p^\text{Ref} + \mathcal{U}(-0.1,0.1)$, $a$ being $x$, $y$, or $z$\\
		$\epsilon_r$ & relative permittivity & $\epsilon_r^\text{Ref} + \epsilon_r^\text{Ref}\mathcal{U}(-0.1,0.1)$\\
		$\sigma$ (S/m) & conductivity & $\sigma^\text{Ref}+\sigma^\text{Ref}\mathcal{U}(-0.1,0.1)$\\
		$c$ (dB) & coupling coefficient & $c^\text{Ref} + \mathcal{U}(-2,2)$\\
		$\abs{E}$ & amplitude of electric field & $\abs{E}^\text{Ref} + \abs{E}^\text{Ref}\mathcal{N}(0,0.025)$\\
		$\angle E$ (radian) & phase angle of electric field & $\angle E^\text{Ref} + \angle E^\text{Ref}\mathcal{N}(0,0.025)$\\
		\bottomrule
	\end{tabular}
\end{table*} 

To simplify the analysis further, the matrix of coupling coefficients with respect to different probes is assumed to be the same. The reference value of coupling coefficients applied is given by
\begin{subequations}
	\begin{equation}
	\mathbf{c}^{x,x} = 
	\begin{bmatrix}
	-0.29-0.27\mathrm{i} & -0.58-0.73\mathrm{i} & -0.46-0.33\mathrm{i} \\
	0.46-1.48\mathrm{i} &  100 & 0.37-1.18\mathrm{i} \\
	-0.33-0.24\mathrm{i} & -0.65-0.83\mathrm{i} & -0.24-0.21\mathrm{i}
	\end{bmatrix}\times 10^{-2},
	\end{equation}
	\begin{equation}
	\mathbf{c}^{x,y} = 
	\begin{bmatrix}   
	-0.3 - 0.3\mathrm{i} & -0.5 - 0.8\mathrm{i} & -0.2 + 0.2\mathrm{i} \\
	1.5 - 7.0\mathrm{i} & 2.5 - 3.8\mathrm{i} & 1.6 - 0.6\mathrm{i}\\
	-0.5 - 1.0\mathrm{i} & -0.5 - 0.7\mathrm{i} & -0.1 + 0.3\mathrm{i}
	\end{bmatrix}\times 10^{-3},
	\end{equation}
\end{subequations} 
and $c_{p_x,p_y}^{y,x}=c_{p_x,-p_y}^{x,y}$, $c_{p_x,p_y}^{y,y}=c_{p_x,-p_y}^{x,x}$ are assumed for the matrices $\mathbf{c}^{y,x}$, $\mathbf{c}^{y,y}$. This assumption is based on the geometric symmetry of the neighboured probes and is valid when the composition of all probes is exactly the same.  Here, the self inductance is not considered and thus the corresponding coefficient is set to 1. Transforming the unit into decibel (\si{\dB}), the uncertainty of the coupling effects is considered by allowing the truly applied coefficient has a deviation up to \SI{2}{\dB}.

\begin{figure}[!h]
	\centering
	\subfloat[$\delta=0$ and $0.01$]{%
		\includegraphics[clip,width=0.6\columnwidth]{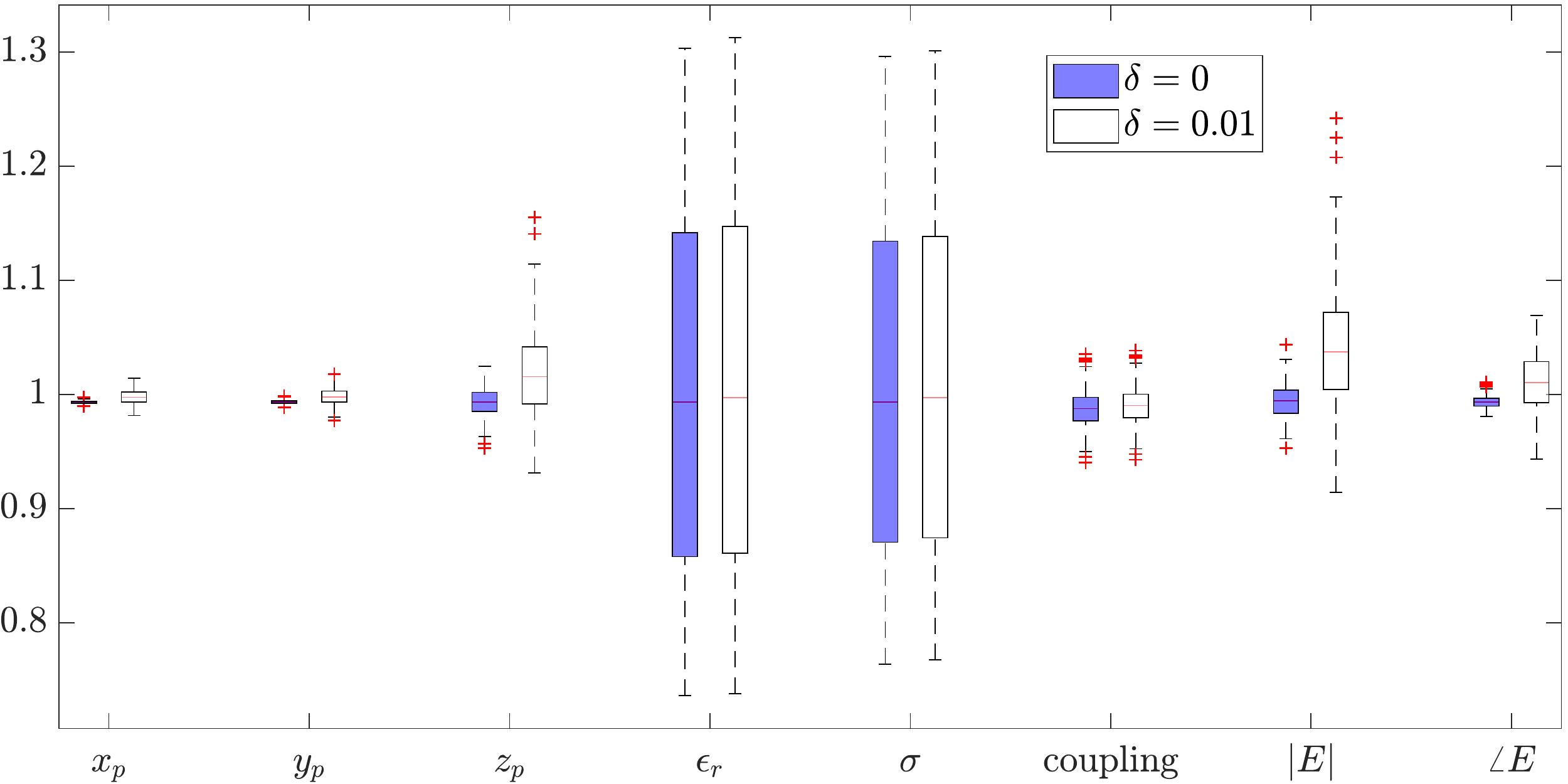}\label{fig:UQ4SourcePercent0_1}
	}\\
	\subfloat[$\delta=0.02$]{%
		\includegraphics[clip,width=0.49\columnwidth]{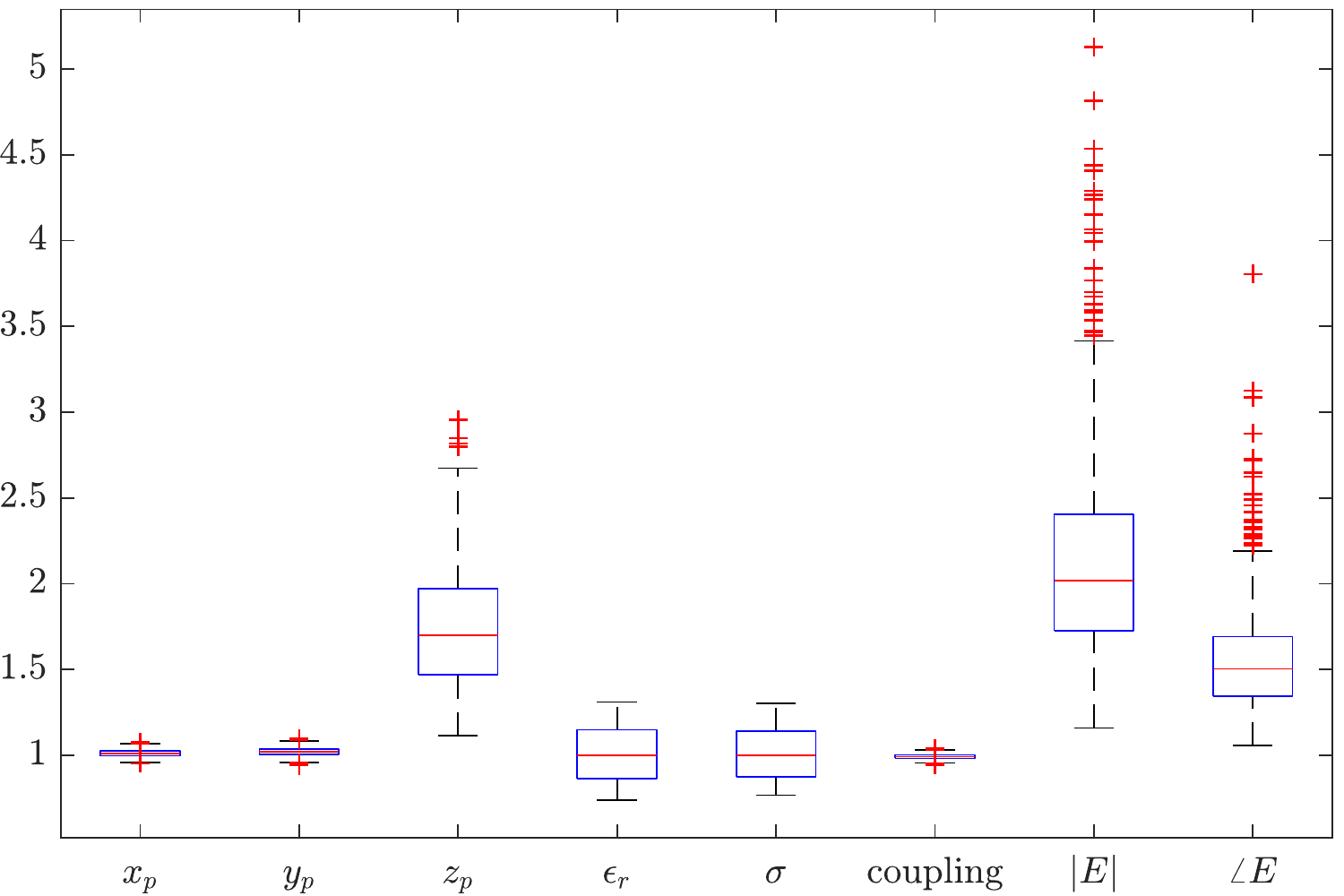}%
		\label{fig:UQ4SourcePercent2}}
	~
	\subfloat[$\delta=0.03$]{%
		\includegraphics[clip,width=0.49\columnwidth]{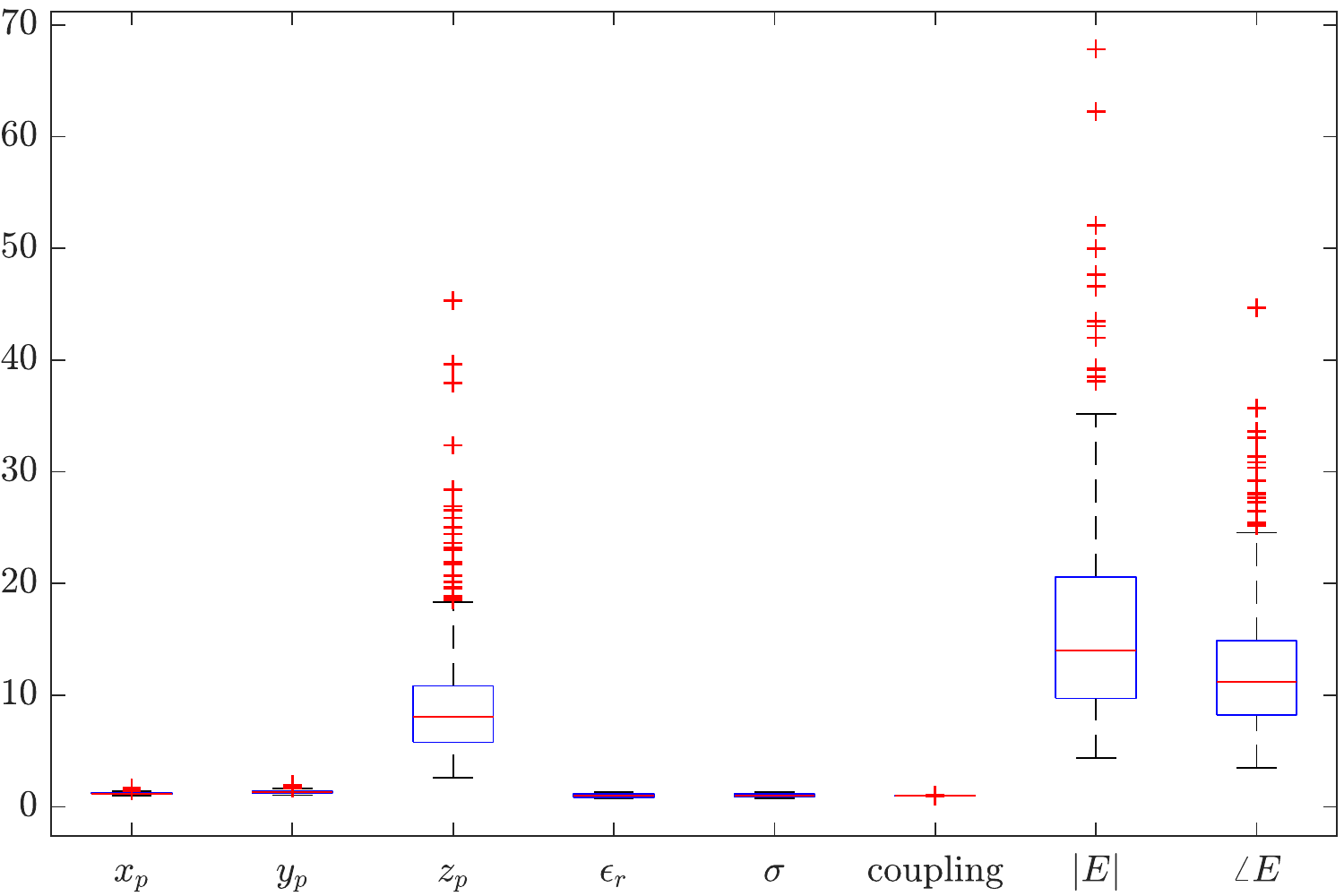}%
		\label{fig:UQ4SourcePercent3}}
	\caption{Box plots of estimated values of 1g SAR based on the reconstructed fields with the PWE approach by setting various values of $\delta$.}
	\label{fig:UQ4Source}
\end{figure}
With $500$ Monte Carlo trials, the estimated \SI{1}{\gram} SAR of the 4th case is plotted in Figure~\ref{fig:UQ4Source} indicating the uncertainty of all factors described in Table \ref{tab:pdf}. As shown in Section \ref{sec:verification},
with no uncertainty, the \SI{1}{\gram} and \SI{10}{\gram} SAR are estimated with a high accuracy for the 4th case. With uncertainties, unbiased estimations are observed when $\delta = 0$, but an overestimation is likely to be obtained when $\delta = 0.01$, 0.02 or 0.03. The extent of the bias increases with $\delta$. Besides, due to the ill-conditioning, the variance of the estimation increases with $\delta$ as higher spatial-frequency spectrum is reconstructed. From the variance, one identifies the most influential factors. When $\delta = 0$ or 0.01, the uncertainty of relative permittivity and conductivity contributes most to the variation of the estimation. Despite the large bias and variance of the estimation, the influence factors are changed as the $z$ coordinate of the probe and the measurement (amplitude and phase) accuracy of the electric field when $\delta = 0.02$ or 0.03. For validation purposes, a further 500 Monte Carlo trials were independently carried out; the obtained results led to the same conclusion.

\subsection{Comparison between the traditional and fast measuring systems}

\begin{figure}[!h]
	\centering
	\includegraphics[width=0.4\linewidth]{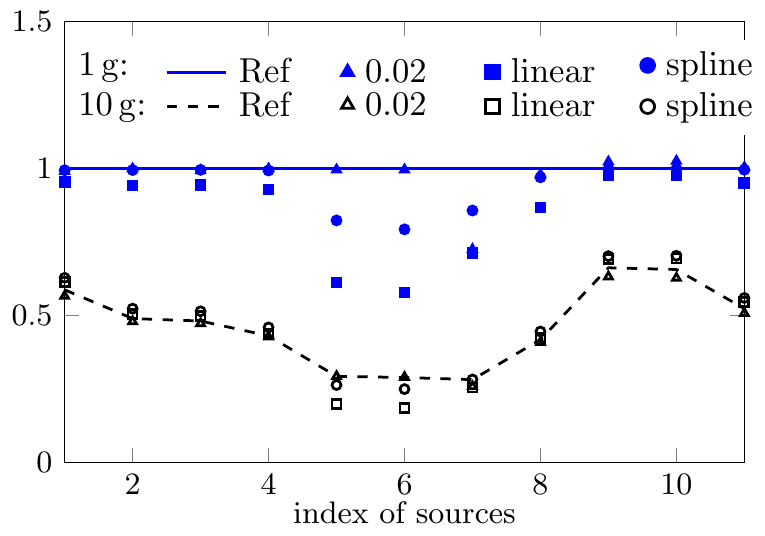}
	\caption{Comparison of estimated SAR by traditional measurement approach and the fast method.}
	\label{fig:compSAR}
\end{figure}
The estimated SAR value of the concerned $11$ cases by the traditional system and the measuring system applying the PWE field reconstructions is presented in Figure~\ref{fig:compSAR}. Both linear and spline interpolation (and extrapolation) are used in the post-processing of the traditional measuring system. While for most cases all approaches yield well approximated estimations, the technique of spline interpolation seems superior to the linear one, most probably because of its greater approximation power. However, for the 5th, 6th, 7th and 8th cases, the estimation by the traditional system is largely biased, especially for the \SI{1}{\gram} SAR. In contrast, although the sampling density is smaller (\SI{7}{\mm} for the fast system and $\le \SI{5}{\mm}$ for the traditional one), the approach based on field reconstructions yields more accurate estimations (except in the 8th case in the estimation of \SI{1}{\gram} SAR). 

\section{Conclusions}
\label{sec:conclusion}
Based on the analytical function and 11 body phantom cases, the traditional and the fast SAR measuring system are studied by simulating the measurement process. While both systems generate well approximated estimations for most of the cases, the approach based on the PWE field reconstruction seems to have the potential to achieve more accurate estimations than the approach based on interpolation and extrapolation. However, the reconstruction algorithm suffers from ill-conditioning, which leads to a trade-off between the reconstruction accuracy and reliability. Reconstructing a high spatial-frequency spectrum gives the possibility of yielding more accurate solutions, but simultaneously the estimation always suffers from a higher variance, and vice versa.
Moreover, the estimation accuracy may vary with emitting sources. Therefore, without the knowledge of the field-reconstruction algorithm and the source, it is hard to conclude which kind of measuring system is superior to the other.  	

Another challenge follows the above conclusion. Due to security and commercial interests, the supplier of a measuring system in general will not provide the code or the details of the reconstruction (for the concerned fast system) or the interpolation (for traditional system) algorithm. As shown by the given results, the traditional system is not the gold standard. Therefore, it is quite necessary to develop a methodology to quantify the measurement accuracy barely based on the product. The involved challenges may include the consideration of various antennas in practice when designing reference antennas, the determination of the reference SAR value, and the uncertainty quantification. 

\section*{Acknowledgments}
This research (Grant number 16NRM07 Vector SAR) has received funding from the EMPIR programme co-financed by the Participating States and from the European Union's Horizon 2020 research and innovation programme. The presented work benefits from the discussion with Dr. Maurício Henrique BEZERRA CARDOSO who did research within project ``Vector SAR" as a postdoctoral researcher at Télécom Paris.

\end{document}